\documentstyle[12pt]{article}
\begin{document} 
\begin{flushright} {OITS 684}\\
December 1999
\end{flushright}
\vspace*{1cm}

\begin{center}  {\large {\bf Erraticity of Rapidity Gaps}}
\vskip .75cm
 {\bf  Rudolph C. Hwa and Qing-hui Zhang}

 {\bf Institute of Theoretical Science and Department of
Physics\\ University of Oregon, Eugene, OR 97403-5203}
\end{center}

\begin{abstract}
The use of rapidity gaps is proposed as a measure of the spatial pattern
of an event.  When the event multiplicity is low, the gaps between
neighboring particles carry far more information about an event than
multiplicity spikes, which may occur very rarely. Two moments of the
gap distrubiton are suggested for characterizing an event.  The
fluctuations of those moments from event to event are then quantified
by an entropy-like measure, which serves to describe erraticity. We
use ECOMB to simulate the exclusive rapidity distribution of each
event, from which the erraticity measures are calculated.  The
dependences of those measures on the order of $q$ of the moments
provide single-parameter characterizations of erraticity.

\end{abstract}

\section{Introduction}

To study the properties of event-to-event fluctuations in multiparticle
production, it is necessary to have an effective measure of the
characteristics of the final state of an event.  The totality of all the
momenta of the produced particles constitutes a pattern.  A useful
measure of a pattern should not contain too much details, but enough to
capture the essence that is likely to fluctuate from event-to-event.  In
previous papers we have used the normalized factorial moments, $F_q$,
as a measure for studying chaos in QCD jets \cite{1}, in classical nonlinear
dynamics \cite{2}, and erraticity in soft production of particles \cite{3}. 
We now consider two different moments in order to improve the analysis
in problems where the use of $F_q$ is less effective. 

The horizontal factorial moments register the multiplicity fluctuation from
bin to bin in an event.  However, when the event multiplicity is low, and
the bin size small, most bins have only one particle per bin, and the
factorial moments fail to provide a good characterization of the event
pattern.  To overcome that deficiency, we shift our emphasis from bin
multiplicities to rapidity gaps.  It is intuitively obvious that the two
quantities are complementary: the former counts how many particles fall
into the same bin, while the latter measures how far apart neighboring
particles are.  Clearly, the former works better when there are many
particles in an event, while the latter is more suitable when there are few
particles.

The search for a good measure of event-to-event fluctuations can be
carried out only if we have an event generator that can be used for the
exploration.  To that end we shall use ECOMB \cite{4}, which simulates soft
production processes in hadronic collision.  It is based on a reasonable
modeling of the many-body dynamics in which the partons undergo
successive color mutation before hadronization.  It is the only model
capable of generating factorial moments that agree with the intermittency
data of NA22 \cite{5}.  However, there is still freedom in the model for
further adjustment.  The aim of this paper is not to test ECOMB or to
improve it.  Despite its imperfections, it can nevertheless simulate events
with sufficient dynamical fluctuations that deviate significantly from
statistical fluctuations.  That capability is what we utilize in our search for
the desired measure.  For that reason it is unnecessary for us to review
here the dynamical content of ECOMB.  After the experimental data are
analyzed and the proposed measure determined, we can then return to
the problem of modeling soft interaction when the new erraticity data will
provide the guidance needed for an upgrading of ECOMB.

\section{The Problem}

Let us start by reviewing the factorial moments for multiparticle
production \cite{6}.  They are defined (for the $q$th order) by
\begin{eqnarray}
f_q=\left<n(n-1) \cdots (n-q+1)\right> \quad ,
\label{1}
\end{eqnarray}
where $n$ is the multiplicity in a bin.  Originally, the average in
(\ref{1}) is performed over all events for a fixed bin, which we now
call vertical average.  Later, the horizontal average is considered for a
fixed event, where $n$ in (\ref{1})  is averaged over all bins.  In either
case if the probability distribution $P_n$ of $n$ can be expressed as a
convolution of the dynamical distribution $D(\nu)$ and the statistical
(Poissonian) distribution, i.e.,
\begin{eqnarray}
P_n = \int d\nu \,\, { 1 \over n!  } \,\, \nu^n \,\, e^{-\nu}\, D(\nu) \quad ,
\label{2}
\end{eqnarray}
then one obtains \cite{6}
\begin{eqnarray}
f_q = \sum^{\infty}_{n=q} { n! \over (n-q)! } P_n = \int d\nu \, \nu^q \,
D(\nu)
\quad .
\label{3}
\end{eqnarray}
Since it is a simple moment of $D(\nu)$, the statistical fluctuation is
regarded as having been filtered out by $f_q$.  
 
The above procedure of eliminating the statistical fluctuation fails either
when the sum in (\ref{3}) does not extend to infinity, or if that fluctuation
cannot be represented by a Poissonian distribution as in (\ref{2}).  Both of
these circumstances occur for horizontal analyses of low multiplicity
events.  There is nothing wrong with calculating $f_q$ according to
(\ref{1}) for such events.  The question is what one can use $f_q$ for.

In \cite{3} the horizontal normalized factorial moments $F_q=f_q/f_1^q$
are used to characterize the spatial pattern of an event.  Such a
characterization clearly cannot convey all the details of an event; indeed,
extensive details using many variables are not desirable for the
quantification of event-to-event fluctuations.  It is evident from (\ref{1})
that only bins with $n\geq q$ can contribute to $f_q$, but the positions of
the contributing bins have no effect on $f_q$.  That deficiency is
unimportant when many bins contribute.  However, when the event
multiplicity $N$ is low and the number of bins $M$ is high, so that the
average bin multiplicity $\bar{n}=N/M\ll 1$, then it is only by large
fluctuations that a bin may have $n\geq q$, whether they are dynamical
or statistical in nature. Since $f_q$ is insensitive to where the few
contributing bins are located, there is very little information about an
event that is registered in $F_q$.  In \cite{7} it is shown that in the
framework of a simple model
$F_q$ are dominated by statistical fluctuations when $N$ is small, but
they reveal the dynamical fluctuations when $N$ is large.

The aim of this paper is to find an alternative to $F_q$ that can effectively
characterize the spatial pattern of an event, even when the event
multiplicity is low.

\section{The Solution}

From (\ref{1}) we see that $f_q$ receives a contribution from a bin in
which $n\geq q$, but ignores where it is located.  In other words $f_q$ is
sensitive to the local height of the rapidity distribution in an event, not to
the spatial arrangement in rapidity.  When $N$ is low and $M$ is high,
many bins are empty.  To have a bin with $n\geq q$ means that even
more bins than average would have to be empty.  It then seems clear that
the complementary information accompanying rapidity spikes is the rapidity
gaps.  When $N$ is high, rapidity gaps are generally not very informative;
however, when $N$ is low, they characterize an event better than counting
spikes.  In the following we shall develop two methods based on
measuring the rapidity gaps.

Since particle momenta can be measured accurately, there is no need to
consider discrete bins in the rapidity space.  Thus we shall work in the
continuum.  Moreover, the advantage of working in the cumulative
variable $X$ has long been recognized \cite{8,9}, and we shall continue to
use the $X$ variable, as in \cite{4} ( though not explicitly stated there in
the first reference).  The definition of $X$ is
\begin{eqnarray}
X (y) = \int^y_{y_{\rm{min}}} \rho(y^\prime) dy^\prime /
\int^{y_{\rm{max}}}_{y_{\rm{min}}} \rho\left(y^\prime\right)
dy^\prime
\quad ,
\label{4}
\end{eqnarray}
where $\rho\left(y^{\prime}\right)$ is the single-particle inclusive
rapidity distribution and
$y_{\rm{min(max)}}$ is the minimum (maximum) value of $y$.  Thus
the accessible range of $y$ is mapped to $X$ between 0 and 1, and the
density of particles in $X$, $dn/dX$, is uniform.

Consider an event with $N$ particles, labeled by $i=1,\cdots,N$, located in
the $X$ space at $X_i$, ordered from the left to the right.  Let us now
define the distance between neighboring particles by
\begin{eqnarray}
x_i = X_{i+1} - X_i , \qquad \qquad  i = 0 , \cdots , N \quad ,
\label{5}
\end{eqnarray}
with $X_0=0$ and $X_{i+1}=1$ being the boundaries of the $X$ space. 
Every event $e$ is thus characterized by a set $S_e$ of $N+1$ number:
$S_e=\left\{x_i|i=0,\cdots,N\right\}$, which clearly satisfy 
\begin{eqnarray} 
\sum^N_{i=0} x_i = 1 \quad .
\label{6}
\end{eqnarray} 
We refer to these numbers loosely as ``rapidity'' gaps.

For any given event $S_e$ contains more information than $P_n$, which
is the bin-multiplicity distribution for that event; in fact, $P_n$ can be
determined from $S_e$, but not in reverse.  To study the fluctuation of
$S_e$ from event-to-event is the most that one can do; indeed, too much
information is conveyed by $S_e$.  For economy and efficiency in
codifying the information we consider moments of $x_i$ that emphasize
large rapidity gaps.  As mentioned earlier, concomitant to the
clustering of particles that results in spikes in the rapidity distribution is
the existence of large gaps.  Thus moments that emphasize large $x_i$
convey similar information about an event as do the moments that
emphasize the high-$n$ tail of $P_n$.  However, the factorial moments of
$P_n$ suffer the defects discussed in Sec. 2 that are absent in the
moments of $x_i$.  The issue of statistical fluctuations has to be
addressed separately.  

Let us then define for each event
\begin{eqnarray}
G_q = { 1 \over N+1 } \sum^N_{i = 0} x^q_i \quad ,
\label{7}
\end{eqnarray}
Despite the similarity in notation, these moments bear no relationship to
the $G$-moments considered earlier \cite{10}.  It is clear from (\ref{6})
and
(\ref{7}) that 
\begin{eqnarray}
G_0 = 1 \qquad {\rm{and}}\qquad G_1 = { 1 \over N+1  } \quad .
\label{8}
\end{eqnarray}
At higher $q$, $G_q$ are progressively smaller, but are increasingly more
dominated by the large $x_i$ components in $S_e$ .  A set of $G_q$ for $q$
ranging up to 5 or 6 is sufficient to characterize an event, better than
$S_e$ itself in the sense that $G_q$ can be compared from event-to-event,
whereas $S_e$ cannot be so compared due to the fluctuations in $N$. 

If we define the gap distribution by
\begin{eqnarray}
g(x) = { 1 \over N+1  } \sum^N_{i = 0} \delta (x-x_i)\quad ,
\label{9}
\end{eqnarray}
then the $G$ moments are
\begin{eqnarray}
 G_q = \int^1_0 dx \, x^q \,  g(x)  \quad .
\label{10}
\end{eqnarray}
This form may become more convenient in some situations.

Since $G_q$ fluctuates from event to event, we can determine
a distribution $P(G_q)$ of $G_q$ after many events.  It is the shape of
$P(G_q)$ that characterizes the nature of
the event-to-event fluctuations of the gap distribution, and therefore of
the spatial pattern of an event.  Again, we can describe $P(G_q)$ by its
moments 
\begin{eqnarray}
C_{p,q} = {  1\over {\cal N} } \sum^{\cal N}_{e = 1} (G^e_q)^p = \int dG_q\,
G^p_q\, P(G_q)
\quad  ,
\label{11}
\end{eqnarray}
where $e$ labels an event and $\cal N$ is the total number of events. 
Since we need not consider bins in $x$, $G_q$ is a number for each event
without statistical error.  Thus calculating the $p$th moment
does not compound statistical errors.  Although one can consider a
range of $p$ moments, we shall focus only on the derivative at $p=1$
in the following.

Since $C_{1,q}=\left<G_q\right>$ is the mean that gives no information on
the degree of fluctuation, the derivative at $p=1$ convey the broadest
information on $P(G_q)$. We have
\begin{eqnarray}
s_q = -{ d \over dp  } C_{p,q} \left.\right |_{p=1} = -\left<{G_q \, ln \,
G_q}\right> \quad  ,
\label{12}
\end{eqnarray}
where $\left<\cdots \right>$ stands for average over all events. The
quantities
$s_q$ are our new measures of erraticity in terms of rapidity gaps.  Since
$G_q$ is not a probability distribution, $s_q$ is not an entropy function,
despite its appearance.

Unlike the factorial moments, $G_q$ does not filter out statistical
fluctuations.  At low multiplicities $F_q$ fails to be effective in that
filtering anyway, as discussed in Sec.\ 2, so it is at no great loss to
consider
$G_q$.  However, we can have an estimate of how much $s_q$ stands out
above the statistical fluctuation by first calculating
 \begin{eqnarray}
s^{st} _q= -\left<G^{st}_q \, ln \, G^{st}_q \right> \quad ,
\label{13}
\end{eqnarray}
where  $G_q^{st}$ is determined from (\ref{10}) by using only the
statistical distribution of the gaps, $g^{st}(x)$, i.e., when all $N$ particles in
an event are distributed randomly in $X$ space.  Then we take the ratio
 \begin{eqnarray}
S_q = s_q / s^{st}_q \quad ,
\label{14}
\end{eqnarray}
and examine how much $S_q$ deviates from 1.  $S_q$ will be the first
erraticity measure that we shall calculate in the next section.

Since our interest is in the deviation of $G_q$ from $\left<G_q \right>$, a
measure of that deviation is 
\begin{eqnarray}
\tilde{s}_q = -\left<{ G_q \over \left<G_q \right>  }  \, ln  \, { G_q \over
\left<G_q \right>  } \right> 
\quad ,
\label{15}
\end{eqnarray}
which clearly would be zero if $G_q$ never deviates from $\left<G_q
\right>$.  We can further normalize $\tilde{s}_q^{st}$ by the
statistical-only contribution
$\tilde{s}_q^{st}$ and define
\begin{eqnarray}
\tilde{S}_q =  \tilde{s}_q / \tilde{s}^{st}_q   \quad .
\label{16}
\end{eqnarray}
Whether this is a better quantity to represent erraticity will be examined
quantitatively in the next section.

The moments $G_q$ are not the only ones that can characterize the
rapidity-gap distribution.  In fact, since $x_i<1$, $G_q$ are usually
$\ll1$, and the statistical errors on $S_q$ and $\tilde{S}_q$ turn out to
be quite large, though not so large as to render the measures
ineffective.  We now consider a different type of moments that also
emphasize the large gaps.  Define for an event with $N$ particles
\begin{eqnarray}
H_q = { 1 \over {N+1}  } \sum^N_{ i=0} (1-x_i)^{-q} \quad ,
\label{17}
\end{eqnarray}
where $x_i$ is as given in (\ref{5}).  These moments also receive
dominant contribution from large $x_i$, as do $G_q$, but $H_q$ can
become $\gg1$.  In terms of $g(x)$ we have 
\begin{eqnarray}
H_q = \int^1_0 \, dx \, (1-x)^{-q} \, g(x)  \quad  ,
\label{18}
\end{eqnarray}
where $g(x)$ must vanish sufficiently fast as $x \rightarrow 1$ to
safeguard the integrability of (\ref{18}).

We can substitute $H_q$ for $G_q$ in all of the foregoing considerations. 
In particular, we can define
\begin{eqnarray}
\sigma_q = \left<H_q \, ln \, H_q \right>\quad ,
\label{19}
\end{eqnarray}
\begin{eqnarray}
\tilde{\sigma}_q = \left<{ H_q \over \left<H_q \right>  }  \, ln \,  { H_q
\over
\left<H_q \right>   } \right>
\quad    ,
\label{20}
\end{eqnarray}
\begin{eqnarray}
\Sigma_q = { \sigma_q \over \sigma^{st}_q  }  \quad \mbox{and} 
\quad     
\tilde{\Sigma}_q = {\tilde{\sigma}_q \over \tilde{\sigma}^{st}_q  }
\quad
\label{21}
\end{eqnarray}
as new measures of erraticity.  The only nontrivial point to remark on
concerns the event average.

For each event $H_q$ depends implicitly on the event multiplicity $N$. 
 If ${\cal P}_n$ is the multiplicity distribution, then the average
$\left<H_q
\right>$ is given by 
\begin{eqnarray}
\left<H_q \right> = \sum^{\infty}_{N = q+1}  H_q(N) {\cal P}_N \quad ,
\label{22}
\end{eqnarray}
where $H_q(N)$ is the mean $H_q$ after averaging over all events with
$N$ particles in each event.  Note that the sum in (\ref{22}) begins at
$N= q+1$, not $0$.  To see this subtle point, let us start with the statistical
average for which we can make precise calculations.  In the Appendix we
show that the probability distribution $p_N(x)$ of the gap distance $x$,
after sampling with sufficiently many events, each with $N$ randomly
distributed particles in the $X$ space, is 
\begin{eqnarray}
p^{st}_N (x) = N \, (1-x)^{N-1} \quad .
\label{23}
\end{eqnarray}
Thus it follows that
\begin{eqnarray}
H^{st}_q (N) = \int^{1}_0 \, dx \, (1-x)
^{-q} \, p^{st}_N(x) = { N \over {N-q}  } \quad .
\label{24}
\end{eqnarray}
Evidently, $N$ must be greater than $q$ to ensure convergence.  If for
statistical calculation we require $n\geq q+1$, then we make the same
requirement for the general problem in (\ref{22}), so that $\sigma_q$
and
$\sigma_q^{st}$ in
(\ref{21}) are calculated on the same basis.

\section{Results}

We have applied ECOMB \cite{4}, upgraded by \cite{11}, to calculate the
rapidity distribution for each event.  From that we compute the gap
distribution $g(x)$ in the $X$ space.  After simulating $10^6$ events at
$\sqrt{s}=20$ GeV, our result for $S_q$ is shown in Fig.\ 1. The
error bars are determined by using the conventional method.  The
straight line in Fig.\ 1 is a linear fit of the central points.  Evidently, the
result indicates a power-law behavior in $q$ for $q\geq 2$
\begin{eqnarray}
S_q \, \propto \, q^\alpha  \quad , \quad     \alpha = 0.156   \quad   .
\label{25}
\end{eqnarray}
The fact that $S_q$ deviates unambiguously from 1 implies that it is a
statistically significant measure of erraticity in multiparticle production. 
At $\sqrt{s}=20$ GeV the average charge multiplicity is only 8.5,
which is low enough to cause problems for the factorial moments
$F_q$, but our use of the gap moments $G_q$ evidently encounters no
similar difficulty.

We next consider $\tilde{S}_q$ defined in (\ref{15}) and (\ref{16}). 
The result is that $\tilde{S}_q$ is nearly independent of $q$, as shown
in Fig.\ 2.  More precisely, we obtain  
\begin{eqnarray}
\tilde{S}_q = 0.96 \pm 0.03 \quad   .
\label{26}
\end{eqnarray}
We regard this result as indicative of the inadequacy of $\tilde{S}_q$ as a
measure of erraticity, since $\tilde{S}_q$ is almost consistent with 1.

Turning to the $H_q$ moments, we show in Fig.\ 3 in semilog plot the
dependence of $\Sigma_q$ on $q$.  Evidently, a very good linear fit is
obtained, yielding
\begin{eqnarray}
 \label{27}
\Sigma_q \, \propto \,  e^{\beta q} \quad , \quad \beta = 0.28
\quad   .
\end{eqnarray}
In the same figure we also show $\tilde{\Sigma}_q$.  Although the
error bars are larger, an exponential behavior
\begin{eqnarray}
\tilde{\Sigma}_q \, \propto \, e^{\tilde{\beta} q} \quad , \quad
\tilde{\beta} = 0.25
\quad ,
\label{28}
\end{eqnarray}
can nevertheless be identified.  Note that  $\tilde{\Sigma}_q$ is much
farther from 1 than $\tilde{S}_q$.  Since $\Sigma_q$ has less statistical
error than $\tilde{\Sigma}_q$, it is more preferred.  Hereafter we shall
discard $\tilde{S}_q$ and $\tilde{\Sigma}_q$ from any further
consideration.  

We now examine the dependence on c.m. energy.  The higher
multiplicities at higher $s$ will decrease the average gap $\left<x \right>$
and the corresponding moments $G_q$ and $H_q$.  We calculate the
effects on
$S_q$ and $\Sigma_q$ at $\sqrt{s}=200$ GeV.  The results are shown in
Figs.\ 4 and 5, where the values for $\sqrt{s}=20$ GeV are reproduced
for comparison.  The power law (\ref{25}) and the exponential behavior
(\ref{27}) persist; the corresponding parameters are
\begin{eqnarray}
\alpha = 0.133 \quad , \quad \beta = 0.108 \quad \mbox{at}
\quad \sqrt{s} = 200 \, \mbox{GeV}
\quad .
\label{29}
\end{eqnarray}
Whereas $\alpha$ has changed little, $\beta$ has decreased
significantly.  The variability of  $\beta$ makes it a more sensitive
measure of erraticity, although the stability of $\alpha$ may
nevertheless be interesting and useful.  Only the analysis of the
experimental data will reveal which one between
$S_q$ and $\Sigma_q$ is better in quantifying erraticity.  It can also
turn out that both are good. 

\section{Conclusion}

We have proposed the moments $G_q$ and $H_q$ as measures of
spatial patterns in terms of rapidity gaps.  We then showed that the
entropy-like quantities $S_q$ and $\Sigma_q$ deviate sufficiently from
1 with small enough statistical errors to serve as effective measures of
erraticity, i.e., event-to-event fluctuations.  In the framework of an soft
hadronic interaction event generator ECOMB we have obtained the
behaviors $S_q \propto q^{\alpha}$ and $\Sigma_q \propto
e^{\beta q}$.  The precise forms of these results are unimportant from
the point of view of the search for an experimental measure to quantify
erraticity.  We offer both $S_q$ and $\Sigma_q$ as our findings.  On
the other hand, from the point of view of using erraticity to test event
generators, then the forms of our results for $S_q$ and $\Sigma_q$ are
pertinent, and the values $\alpha = 0.156$ and $\beta = 0.28$ are
useful for comparison with the soft production data.  Analysis of the
data, especially those of NA22, to determine $S_q$ and $\Sigma_q$ is
therefore urged.  The experimental values of $\alpha$ and $\beta$
will either eliminate wrong models or provide crucial guidance to the
improvement of the correct models.

The extension of this approach to other collision processes is obviously the
next step.  For heavy-ion collisions the multiplicities will be too high for
any interesting study in rapidity gaps, unless one focuses on more rarely
produced particles, such as $J/\psi$.  At RHIC when only $pp$ collisions
are studied, our result for $\sqrt{s}=200$ GeV can be tested.  For
nuclear collisions very narrow $\Delta p_T$ selection must be made to
render the rapidity gap analysis meaningful.

A natural direction of generalization is, of course, to higher
dimensional analysis.  One-dimensional gaps should be generalized to
two-dimensional voids, which is more difficult to define if the use of
bins is to be avoided.  When a good measure is found, not only can it
be employed as an alternative to the multiplicity analysis in the lego
plot, useful application can no doubt be found also in the study of
galactic clustering in astrophysics.  Finally, in view of the abundance of
experimental data and the variety of event generators for $e^+ e^-$
annihilation, a generalization to the multidimensional variables
suitable for such problems will be a fruitful direction to pursue.

\begin{center}
\subsubsection*{Acknowledgment}
\end{center}

We are grateful to Dr.\ Z.\ Cao for helpful communication at the
beginning of this research project.  This work was supported in part by
U.\ S.\ Department of Energy under Grant No.\ DE-FG03-96ER40972.

\newpage
\begin{center}
\subsection*{Appendix}
\end{center}

We derive in this Appendix the probability distribution of gaps in the
purely statistical case.

The gap distribution $g(x)$ defined in (\ref{9}) and (\ref{5}) can be
more elaborately written as $g_e (x; X_1, \cdots, X_N)$ for the $e$th
event with
$N$ particles located at $X_1, \cdots, X_N$.  It has the normalization
\begin{eqnarray}
\int^{1}_0 dx \, g_e (x; X_1, \cdots, X_N) = 1 \quad   .
\label{A1}
\end{eqnarray}
If the value of $X_i$ is randomly selected in the interval $0 \leq X_i \leq
1$, then after a large number of events ${\cal N}$ with $N$ particles
in each, probability distribution in $x$ is
\begin{eqnarray}
p^{st}_N(x) &=& {  1\over \cal N  } \sum^{\cal N}_{e=1} g_e (x; X_1,
\cdots, X_N)\nonumber\\ 
&=& N! \int^{1}_0 dX_1 \int^{1}_{X_1} dX_2
\cdots
\int^{1}_{x_{N-1}} dX_N\ g_e (x; X_1, \cdots, X_N) \quad ,
\label{A2}
\end{eqnarray}
 in which the primitive distributions of the individual
$X_i$ values that should appear inside the integral have been set equal
to 1 due to the statistical nature of their occurrences.

Consider a specific gap $y = X_j - X_i$, where $j = i+1$.  Then for all the
multiple integrals at and before $X_i$, we may reverse the order of
integration and obtain
\begin{eqnarray}
 \int^{1}_0 dX_1 \int^{1}_{X_1} dX_2 \cdots \int^{1}_{X_{i-1}} dX_i =
\int^{1}_0 dX_i \int^{X_i}_0  dX_{i-1} \cdots \int^{X_2}_0 dX_1 =
\int^{1}_0 dX_i { 1 \over (i-1)!  } X^{i-1}_i  \, .
\label {A3}
\end{eqnarray}
 For all the multiple integrals at and after
$X_j$, we have 
\begin{eqnarray}
\int^{1}_{X_j} dX_{j+1} \cdots \int^{1}_{X_{N-1}} dX_N = { 1 \over (N-j)!  }
(1-X_j)^{N-j}  \quad   .
\label {A4}
\end{eqnarray}
 Substituting these and (\ref{5}) and (\ref{9}) into (\ref{A2}), we get 
\begin{eqnarray}
p^{st}_N(x) &=& { N! \over N+1  } \sum^{N}_{i=0} \int^{1}_0 dX_i
\int^{X_i}_0 dX_j  { {X_i}^{i-1} \over (i-1)!  } \quad { (1-X_j)^{N-j} \over
(N-j)!  } \,\,
\delta (x-X_j+X_i)\nonumber\\
& = &{ N! \over N+1  } \sum^{N}_{i=0}
\int^{1}_0 dy
\int^{1-y}_0 dX_i \,\, { X^{i-1}_i \over (i-1)!  } \quad { (1-X_i+y)^{N-j}
\over (N-j)!  } \,\, \delta(x-y)  \quad.
\label{A5}
\end{eqnarray}
The integral over
$X_i$ yields $(1-y)^{N-1}/(N-1)!$. Thus the final result is 
\begin{eqnarray}
p^{st}_N(x) = N(1-x)^{N-1}   \quad  .
\end{eqnarray}
This
behavior is verified by numerical simulation.

\vspace{2cm}

\section*{Figure Captions}

\begin{description}
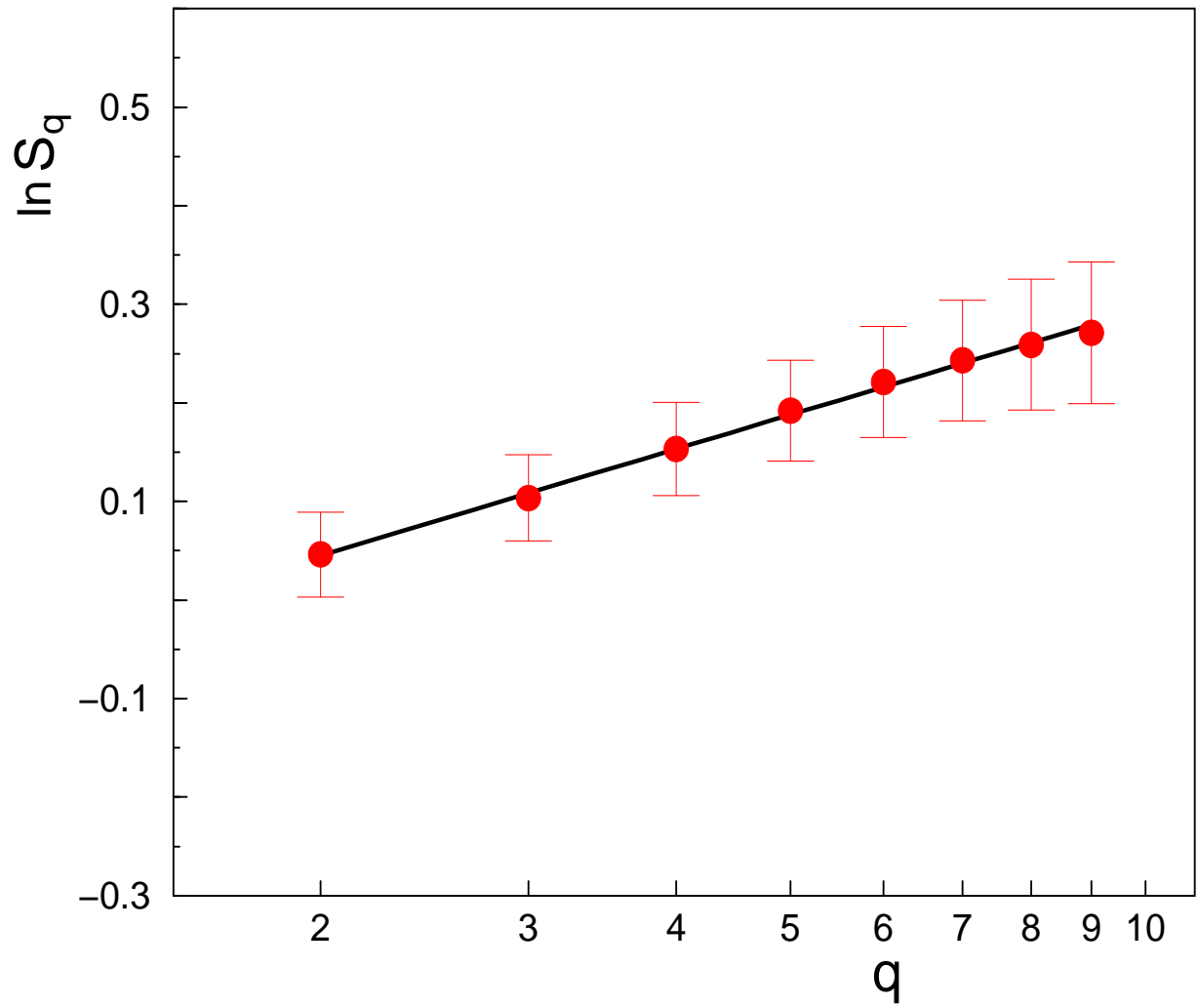
\item[Fig.\ 1]$\ell n S_q$ vs $q$ as determined by ECOMB.  The solid
line is the best fit of the central points.
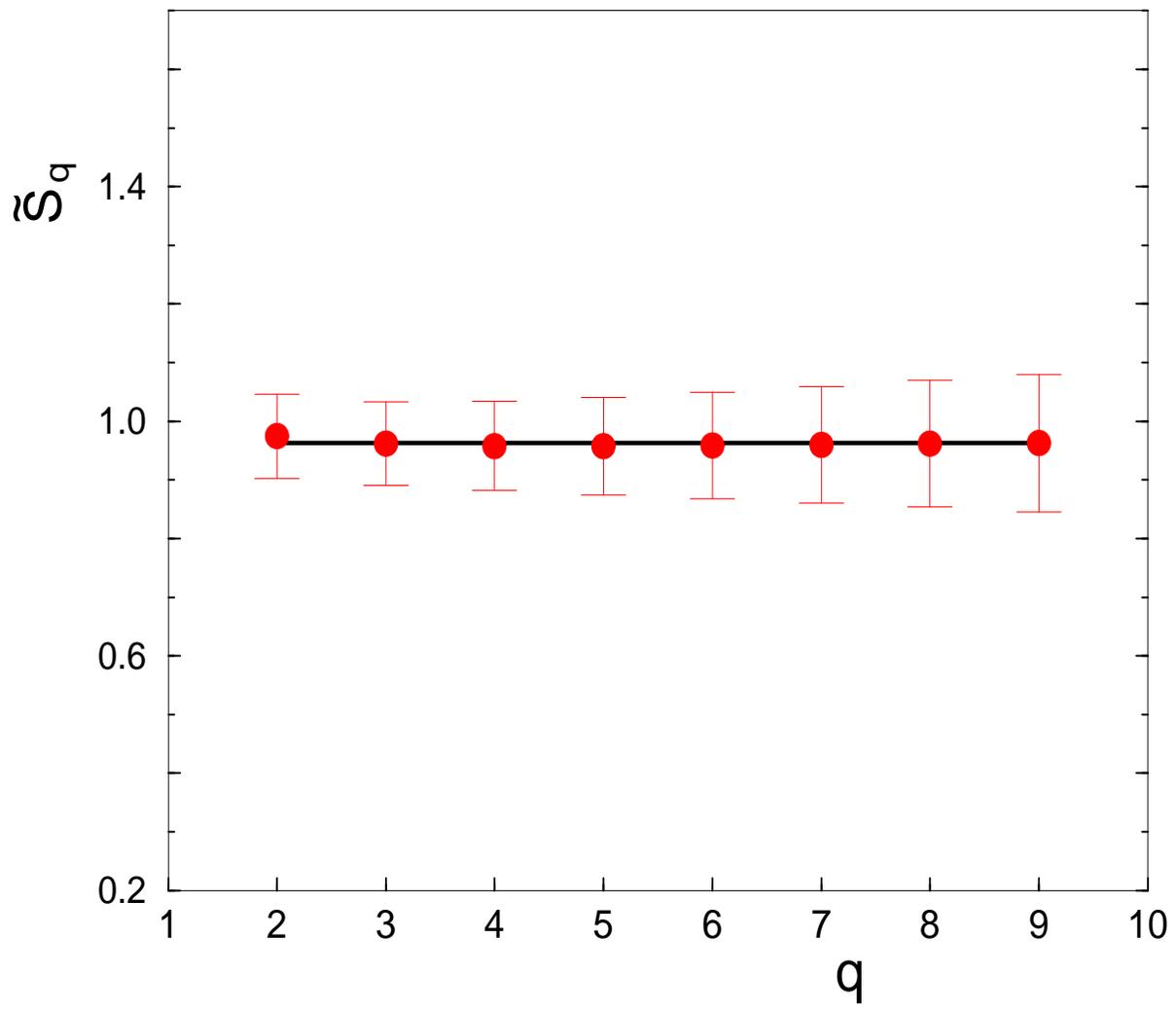
\item[Fig.\ 2]$\tilde{S}_q$ vs $q$ with the same comments as in Fig.\ 1.
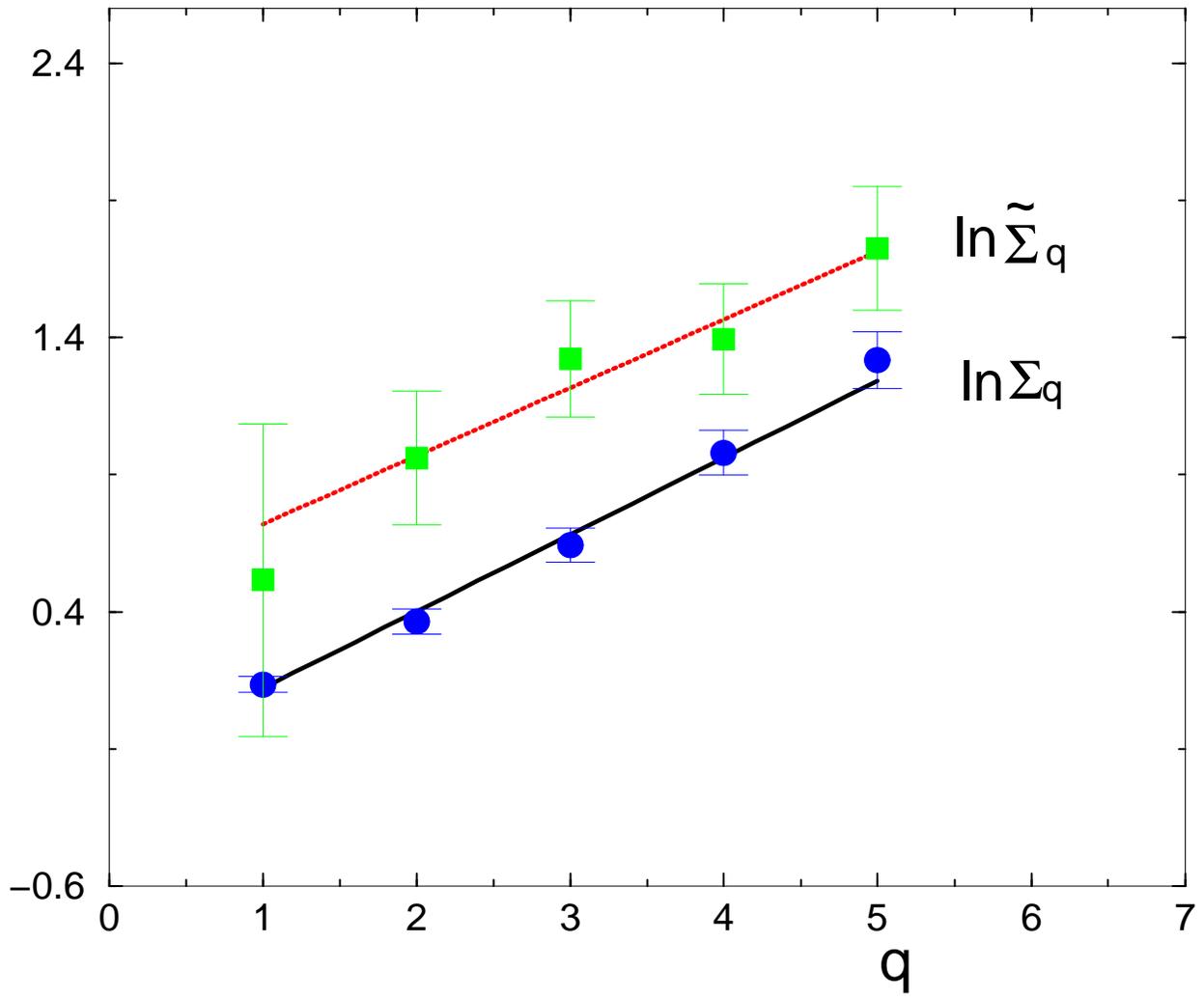
\item[Fig.\ 3]$\ell n \Sigma_q$ and $\ell n \tilde{\Sigma}_q$ vs $q$
with the same comments as in Fig.\ 1.
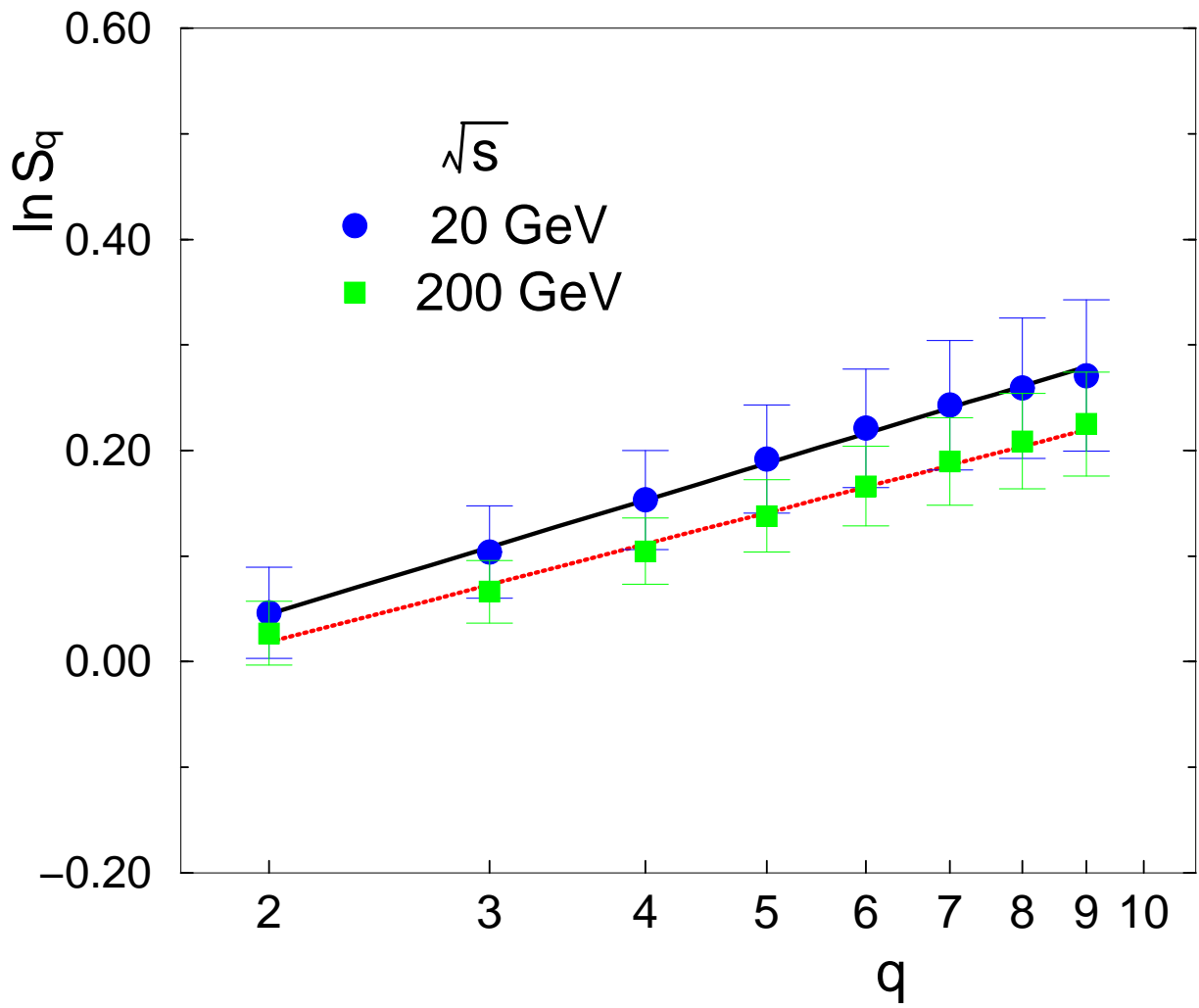
\item[Fig.\ 4]$\ell n S_q$ vs $q$ at two different energies.
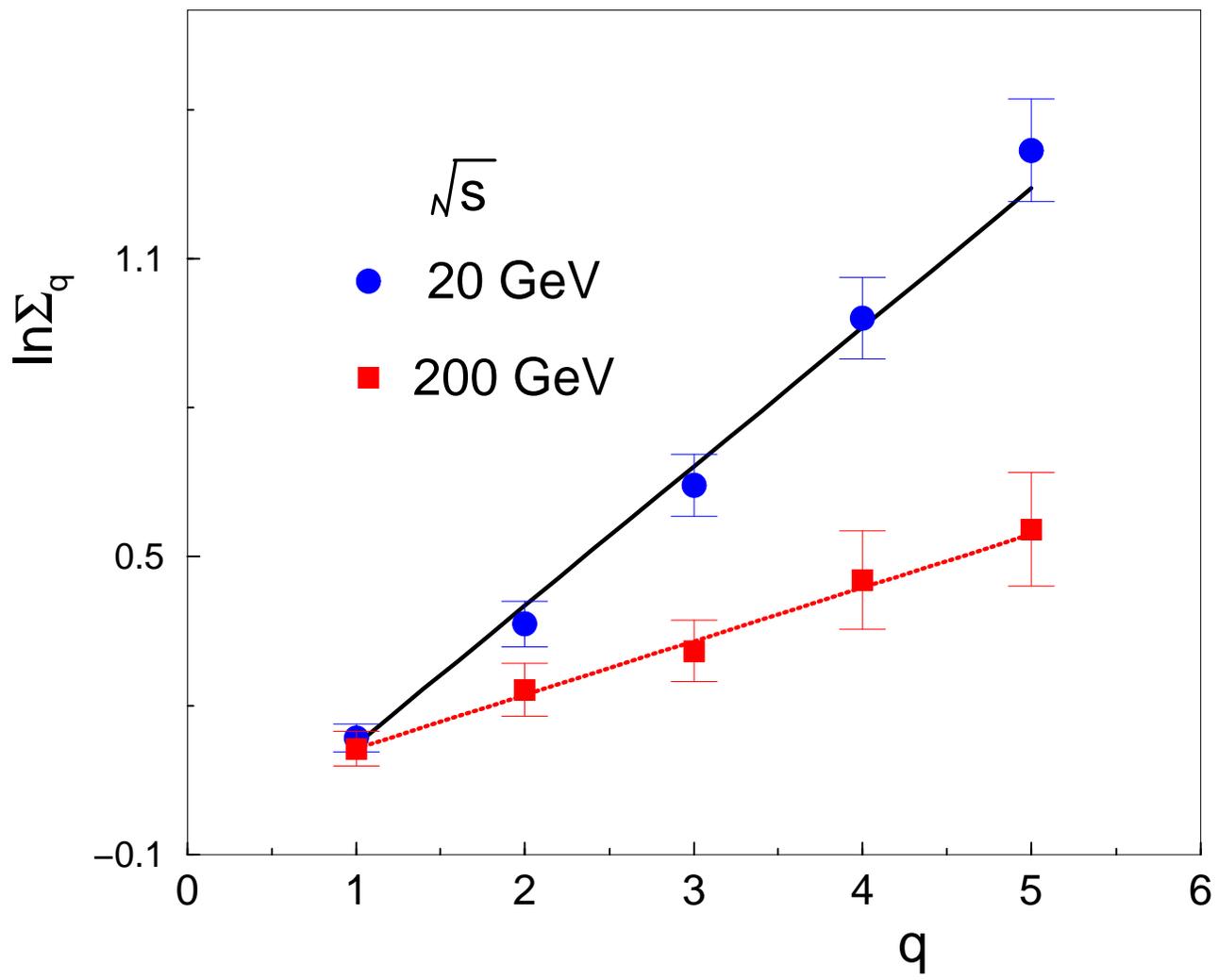
\item[Fig.\ 5]$\ell n \Sigma_q$ vs $q$ at two different energies.

\end{description}


\newpage
\begin{thebibliography}{000}

\bibitem{1} 
Z.\ Cao and R.\ C.\ Hwa, Phys.\ Rev.\ Lett.\ {\bf 75},
1268 (1995); Phys.\ Rev.\ D {\bf 53}, 6608 (1996);  {\it ibid} {\bf 54},
6674 (1996).

\bibitem{2}Z.\ Cao and R.\ C.\ Hwa, Phys.\ Rev.\ E {\bf 56}, 326
(1997).

\bibitem{3} Z.\ Cao and R.\ C.\ Hwa, hep-ph/9901256,  Phys.\ Rev.\
D (to be published).

\bibitem{4}Z.\ Cao and R.\ C.\ Hwa, Phys.\ Rev.\ D {\bf 59}, 114023
(1999).

\bibitem{5} I.\ V.\ Ajineko {\it et al}.\ (NA22), Phys.\ Lett.\ B {\bf
222}, 306 (1989); {\bf 235}, 373 (1990).

\bibitem{6} A.\ Bia\l as and R.\ Peschanski, Nucl.\ Phys.\ B {\bf
273}, 703 (1986); {\bf 308}, 857 (1988).

\bibitem{7}J.\ Fu, Y.\ Wu, and L.\ Liu, hep-ph/9903217

\bibitem{8}A.\ Bia\l as and M.\ Gardzicki, Phys.\ Lett.\ B {\bf 252},
483 (1990)

\bibitem{9} E.\ A.\ DeWolf, I.\ M.\ Dremin, and W.\ Kittel, Phys.\ Rep.
{\bf 270}, 1 (1996).

\bibitem{10} R.\ C.\ Hwa, Phys.\ Rev.\ D {\bf 41}, 1456 (1990); I.\
Derado, R.\ C.\ Hwa, G.\ Jancso, and N.\ Schmitz, Phys.\ Lett.\ B {\bf
283}, 151 (1992).

\bibitem{11}R.\ C.\ Hwa and Y.\ Wu, Phys.\ Rev.\ D {\bf 60}, 097501
(1999).

\end{thebibliography}
\end{document}